\documentstyle[12pt]{article}

\def\n{\nu}
\def\l{\lambda}

\begin{document}

\def\singlespace 
{\smallskipamount=3.75pt plus1pt minus1pt
\medskipamount=7.5pt plus2pt minus2pt
\bigskipamount=15pa plus4pa minus4pt \normalbaselineskip=12pt plus0pt
minus0pt \normallineskip=1pt \normallineskiplimit=0pt \jot=3.75pt
{\def\smallskip {\vskip\smallskipamount}} {\def\medskip
{\vskip\medskipamount}} {\def\bigskip {\vskip\bigskipamount}}
{\setbox\strutbox=\hbox{\vrule height10.5pt depth4.5pt width 0pt}}
\parskip 7.5pt \normalbaselines} 

\def\middlespace
{\smallskipamount=5.625pt plus1.5pt minus1.5pt \medskipamount=11.25pt
plus3pt minus3pt \bigskipamount=22.5pt plus6pt minus6pt
\normalbaselineskip=22.5pt plus0pt minus0pt \normallineskip=1pt
\normallineskiplimit=0pt \jot=5.625pt {\def\smallskip
{\vskip\smallskipamount}} {\def\medskip {\vskip\medskipamount}}
{\def\bigskip {\vskip\bigskipamount}} {\setbox\strutbox=\hbox{\vrule
height15.75pt depth6.75pt width 0pt}} \parskip 11.25pt
\normalbaselines} 

\def\doublespace 
{\smallskipamount=7.5pt plus2pt minus2pt \medskipamount=15pt plus4pt
minus4pt \bigskipamount=30pt plus8pt minus8pt \normalbaselineskip=30pt
plus0pt minus0pt
\normallineskip=2pt \normallineskiplimit=0pt \jot=7.5pt
{\def\smallskip {\vskip\smallskipamount}} {\def\medskip
{\vskip\medskipamount}} {\def\bigskip {\vskip\bigskipamount}}
{\setbox\strutbox=\hbox{\vrule height21.0pt depth9.0pt width 0pt}}
\parskip 15.0pt \normalbaselines}

\def\l{\lambda}
\def\lp{\lambda^\prime}
\def\lpp{\lambda^{\prime \prime}}
\def\lbf{\lambda^{\prime \prime}_{133}}
\def\lbs{\lambda^{\prime \prime}_{233}}
\def\lbv{\lambda^{\prime \prime}_{B\!\!\!/}}
\def\lbi{\lambda^{\prime \prime}_{i33}}
\def\abv{A^0_{B\!\!\!/}}
\newcommand{\be}{\begin{equation}}
\newcommand{\ee}{\end{equation}}
\newcommand{\bea}{\begin{eqnarray}}
\newcommand{\eea}{\end{eqnarray}}
\begin{center}
\large {\bf Antisymmetric neutrino Yukawa matrices and neutrino mixing} \\ 
\vskip 1in
Biswajoy Brahmachari \\
\end{center}
\begin{center}
Department of Physics and Astronomy,\\
University of Maryland, College Park,\\
MD-20740, USA.
\\
\end{center}
\vskip 1in
{
\begin{center}
\underbar{Abstract} \\
\end{center}
{ {\small                                                                    
We study leptonic CKM mixing matrices when the neutrino Yukawa matrices 
are antisymmetric which gives rise to mass patterns suitable to explain 
solar, atmospheric and LSND neutrino oscillation experiments. Taking 
diagonal leptonic matrices which can give rise to hierarchical lepton 
masses, we compute the leptonic CKM matrix.   

}
}
\newpage

\section{Introduction}

The structure of the Yukawa coupling matrices in the generation space are
left unconstrained by gauge symmetries which can experimentally be
tracked by measuring the masses and the mixing angles of the 
fermions\cite{pdg}. Although the masses (eigenvalues) of the quarks and the 
charged leptons are hierarchical in pattern, the solar\cite{lang} and 
atmospheric\cite{atm} neutrino oscillation results, if caused by 
$\n_e \leftrightarrow \n_\mu$ and $\n_\mu \leftrightarrow \n_\tau$ 
oscillations, hint towards an approximate degeneracy of squared neutrino 
masses in the three generation scenario whether in the case of four 
neutrino species including a sterile neutrino $\n_s$ the approximate 
degeneracy is at least partial\cite{schramm}; In this case 
the LSND result on $\n_e \leftrightarrow \n_\mu$ indicating $\Delta 
m^2_{e\mu} \sim 1$ eV$^2$ can also be accommodated. It is worth noting 
that whereas the quark mixing angle $V_{cb}$ is approximately 0.04 the 
above explanation of the atmospheric neutrino anomaly requires that the 
corresponding mixing angle in the leptonic sector is maximal, requiring 
$\sin^2 2\Theta_{\mu s} \sim O(1)$. Mass matrices of the quark and 
leptonic sectors can indeed be very different. Because the Yukawa 
coupling matrices are independent of the gauge symmetries, and also even 
though the underlying GUT symmetries\cite{leem,bm} may relate various 
Yukawa matrices they do not describe the form of the Yukawa matrices, we 
are motivated to study the leading symmetry properties of the Yukawa 
matrices\cite{tex1,tex2} themselves vis-a-vis the experimental 
informations on fermion masses, in particular the neutrinos. 
\section{Completely antisymmetric masses}
Let us consider that an $n$ dimensional antisymmetric matrix M has an 
eigenvalue $\lambda_0$. Then we have, 
\be
{\bf \rm Det}[-{\bf 1} \lambda_0 - M] = (-)^n~{\bf \rm Det}[ {\bf 1} 
\lambda_0 - M]=0 \label{e1} 
\ee 
implying that $-\lambda_0$ is also an eigenvalue. Thus a $3 \times 3$ 
antisymmetric matrix must have a zero eigenvalue. However, as noted above 
if the apparent solar neutrino deficit has to be explained in terms of 
matter induced resonant Mikhayev-Smirnov-Wolfenstein (MSW) or vacuum 
two-flavor oscillations and if the atmospheric neutrino anomaly is due to 
$\nu_\mu \leftrightarrow \nu_\tau$ oscillations the three neutrino 
species have to be almost degenerate in mass. Thus, within an 
antisymmetric scenario all three neutrinos must have almost vanishing 
mass ruling out neutrino masses in the eV range. Thus it is difficult to 
consider neutrino as a hot dark matter candidate. Therefore we consider 
the following mass matrix of the four neutrino species including a 
sterile neutrino, 
\be
M=\pmatrix{0 & m_{12} & m_{13} & m_{14} \cr
           -m_{12} & 0 & m_{23} & m_{24} \cr
           -m_{13} & -m_{23} & 0 & m_{34} \cr
           -m_{14} & -m_{24} & -m_{34} & 0}. \label{e2}
\ee
The eigenvalues of M satisfies the equation,
\be
 \l^4+(m^2_{12}+m^2_{13}+m^2_{14}+m^2_{23}+m^2_{24}+m^2_{34})~\l^2+
(m_{14}m_{23}-m_{13}m_{24}+m_{12}m_{34})^2=0.
\ee
Firstly, the equation is invariant under $\l \leftrightarrow -\l$ 
in accordance with Eqn (\ref{e1}). However, in this case we have two 
non-zero solutions of $\l^2$. Hence, four solutions are grouped into two 
sets, $\{\n_e,\n_\tau\}$ and $\{\n_\mu,\n_s\}$ \footnote{At this stage we 
could have also chosen the pairs as $\{\n_e,\n_s\}$ and 
$\{\n_\tau,\n_\mu\}$, because they are indistinguishable from the point of 
view of the mass matrix as long as antisymmetry is unbroken. At later 
stage this choice will be justified by the mixing angles favoured by 
experiments.}. Each set having a pair of eigenvalues with equal magnitude 
and opposite sign as a result of the antisymmetry independent of the 
entries of the matrix! This guarantees a mass squared degeneracy among 
each set, hence solar neutrino problem can in principle be described by 
$\n_e \leftrightarrow \n_\tau$ oscillations and atmospheric neutrino 
problem by the $\n_\mu \leftrightarrow \n_s$ oscillations. Moreover 
$\Delta m^2_{e \mu}$ is of the order of $m^2_{ij}$ where $m_{ij}$ is a 
typical entry of the 
matrix $M^D_\n$. Choosing $m_{ij}~1$ eV we can get the mass squared 
difference required by the LSND result, and as a bonus 
$\sum_{i=1}^{4}|\l_{i}| \sim$ O(1) eV, making neutrino favorable as a 
Dark Matter candidate.   
\section{See-saw mechanism}
In the discussion above we did not make an attempt to explain the 
smallness of the entries in $M^D_\n$, and indeed the absence of 
Majorana mass terms forbid see-saw mechanism\cite{seesaw} forbidding a 
natural way to explain the smallness of neutrino mass. Now we present a 
modified model to include Majorana terms and thereby see-saw 
mechanism. The skeleton key to the following discussion is that the 
eigenvalues of a matrix $MM^\dagger$ are the squares of those of $M$. The 
see-saw mechanism suppresses the Dirac mass term and we get back a 
light left handed Majorana neutrino mass matrix, 
\be
M^\n_{ij}= ({V^2_F \over 2~V_R })~({\tan^2 \beta 
\over 1 + \tan^2 \beta}) ~\epsilon_{ik} ~[\chi]^{-1}_{kl} 
~\epsilon^{\dagger}_{lj} \label{e6} 
\ee
Where $\tan \beta \equiv {<H_u> \over <H_d>}$ and $\chi$ is the 
right handed Majorana type Yukawa texture. We note that if $\chi$ is an 
approximately diagonal matrix, the light neutrino mass eigenvalues keep 
the underlying pattern dictated by those of the Dirac mass 
textures $\epsilon$. 
\section{Four neutrino textures}
Here we postulate a $4 \times 4$ Dirac mass textures as, 
\be
M^D_\n=
  ({V_F \over \sqrt{2}})~({\tan^2 \beta          
\over 1 + \tan^2 \beta})~\pmatrix{0 & 1 & 1 & 1 \cr
                                  -1 & 0 & 1 & 1 \cr
                                  -1 & -1 & 0 & 1 \cr
                                  -1 & -1 & -1 & 0 } 
\longleftarrow{{ \rm DIRAC~TEXTURE}}\label{e7}
\ee
The Majorana Yukawa texture is defined as, 
\be
M_R = V_R~~
\bordermatrix
{& \n^e_{R} & \n^\tau_R & \n^\mu_{R} & \n^s_R \cr
\hline 
&&&&\cr
\overline{({\n^e_{R}})^c}    & 1 & 0 & 0 & 0 \cr
 \overline{({\n^\tau_{R}})^c}    & 0 & 1 & 0 & 0 \cr
 \overline{({\n^\mu_{R}})^c}  & 0 & 0 & 1 & 0 \cr
 \overline{({\n^s_{R}})^c} & 0 & 0 & 0 & 1+ \eta \cr
}
\longleftarrow {\rm MAJORANA~TEXTURE} 
\label{e8} \ee 
Where $\eta~V_R$ is the mass of the sterile neutrino. We expect it to be 
a little different from the other right handed masses. Using the 
expression for the Dirac and Majorana masses given in Eqn(\ref{e7}) and 
Eqn.(\ref{e8}) and inserting them to Eqn(\ref{e6}) we can describe the $4 
\times 4$ light neutrino Majorana mass matrix in terms of three parameters 
$V_R$, $\tan~\beta$ and $\eta$\footnote{For renormalization group 
analysis of the scale $V_R$ in the presence of right handed triplet scalars 
see Ref\cite{rgelr}.}. The dependence of the results on 
$\tan~\beta$ is milder than that on $\eta$, and $V_R$ is restricted from 
unification requirements. Our results are summarized in Table 1.

\section{Solar, atmospheric and LSND neutrino oscillations}

As we have noted above, if the apparent solar neutrino deficit has to be 
explained in terms of matter induced resonant Mikhayev-Smirnov-Wolfenstein 
(MSW) or vacuum two-flavor oscillations, the most recent `best fit' mass 
squared differences for $\delta m^2_{e\mu}$ or $\Delta m^2_{e\tau}$ is in 
the range $10^{-5}$ and also, if the atmospheric neutrino anomaly is due to 
$\nu_\mu \leftrightarrow \nu_\tau~{\rm or}~\nu_s$\cite{liu}oscillations, we need 
$\delta m^2_{\mu\tau/s} \sim 10^{-2}-10^{-4} {\it eV^2}$ with large mixing. 
Moreover LSND oscillation requires $\Delta m^2_{e \mu} > 0.3$ eV$^2$. Our 
textures can predict $\Delta m^2_{e\tau}=10^{-5}$ eV$^2$ and $\Delta 
m^2_{\mu s}=10^{-2}$ eV$^2$ and $\Delta m^2_{e\mu} = 1$ eV$^2$ for ranges of 
parameters as displayed in table 1. We note that the corresponding mixing 
angles are not solely the properties of the neutrino mass matrices, as 
neutrino mixing has to be folded in with the mixing of the charged 
leptons, whereas the required mass squared differences 
depend exclusively on the mass matrices. However, to convince ourselves 
that there exists compatible mixing 
angles we assume that the leptonic mass matrix is diagonal which makes 
the leptonic mixing matrix identical to the neutrino mixing matrix. 
Thereafter, we parametrize the $4\times 4$ light neutrino mixing 
matrix according to the convention of Barger et al\cite{barger}, which gives
$\sin^2 2\Theta_{e \tau} \approx 0.75$, $\sin^2 2\Theta_{e \mu} \approx 
0.88$ and $\sin^2 2 \Theta_{\mu s} \approx 0.9$\footnote{A $4\times4$ 
unitary matrix can be parametrized by six angles and two phases. In our 
case the rotation matrix is real. Out of the three remaining angles two 
are large and one vanish.}: large angles for all the 
transitions\footnote{Zenith angle distribution of atmospheric sub GeV 
data favours large angle solution of the solar neutrino 
anomaly\cite{teshima}, at least in the three flavour case.} under 
consideration. These mixing angles can be altered by 
choices of the Yukawa coupling matrix of the charged leptons which is 
beyond the scope of the present discussion.

\section{Constraints from double beta decay} 

Neutrinoless double beta decay is unobserved in nature. The 
Heidelberg-Moscow Experiments quote\cite{klapdor} the lower limits on the 
half life as $\Gamma_{1/2} \ge 1.1~10^{25}$ y; which already restricts 
the $<\n_e\n_e>$ Majorana mass term to be less that 0.60 eV at 90 $\%$ 
confidence level\footnote{The limit depends on nuclear matrix elements. 
See Table (4) of Ref\cite{klapdor}.}. In future the limits may go down to 
0.1 eV with the present experimental setup. Thus in combination with LSND 
results, which require $\Delta m^2_{e \mu} \approx 0.3$ eV$^2$ or higher 
(a lower limit on mass$^2$ implys an upper limit on $V_R$), neutrino-less 
double beta decay constraints (vice-versa) may establish or rule out 
these textures. 

\section{Hot Dark Matter} 

$\Omega=1$ with $h \approx 0.5$ Cold Dark Matter cosmological models 
provide a good fit to the observational data in the presence of massive 
neutrinos, when $m_\n \approx 5 eV$ is equally shared between two relatively 
heavy neutrinos, contributing a tiny Hot component (CHDM) to the dark 
matter\cite{primack}. Our textures do not achieve this as that will give 
too much Majorana mass to the electron neutrino. At best these textures 
can give a pair of neutrinos approximately at 1 eV. On the contrary, 
neutrinos in this mass range might modify the power spectrum to 
agree better with the data on galaxy distribution in the $\Omega \approx 
0.4$, $\Omega_\Lambda \approx 0.6$ cosmology indicated by the 
high-red-shift supernova data\cite{primack1}, which may resolve the 
problems in r-process nucleosynthesis\cite{fuller} in Type II supernovae. 

\section{Neucleosynthesis bounds}

There exist bounds on the product $\Delta~m^2_{is}~\sin^4~2~\Theta_{is}$ 
form Big bang nucleosynthesis. These bounds are derived by demanding 
that oscillations do not bring the sterile neutrino in equilibrium 
with the known neutrino species. However, these bounds 
depends on the value of the primordial lepton asymmetry. If the 
primordial asymmetry is of the order of $10^{-9}$ and the mixing is 
maximal, the bound upper bound from neucleosynthesis is nearly $\delta~m^2 
\approx 10^{-8}\cite{bbn1}$. However, if the initial lepton asymmetry is 
large enough $L_e \approx 10^{-5}$ the bound on the mixing between an 
ordinary and sterile neutrino is weakened and large-angle-mixing solution 
of the atmospheric neutrino anomaly becomes 
feasible\cite{bbn2}. 
\section{Theoretical scenario}
It is well known that a straight-forward generalization of three generation 
scenario of the Dirac sector to four generations is problematic from the 
point of view of the invisible Z width which implys that only three light 
neutrino species couple to the Z bosons. This experimental restriction 
can be circumvented by adding a pair of singlet sterile neutrinos 
$\n^s_L$ and $\n^s_R$ to the three generation of quarks and leptons. Let 
us write down the neutrino Dirac type mass terms in a matrix form, 
\be
 \cal{L^D} =
\bordermatrix
{& \n^e_{R} & \n^\mu_R & \n^\tau_{R} & \n^s_R \cr
 \n^e_{L} & 0 & H_2 & H_2 & H_2 \cr
 \n^\mu_{L} & -H_2 & 0 & H_2 & H_2 \cr
 \n^\tau_{L} & -H_2 & -H_2 & 0 & H_2 \cr
 \n^s_{L} & -H_s & -H_s & -H_s & 0 \cr
}
\ee
Where the Yukawa couplings could be either zero or one, and the mass 
spectrum is generated due to the combination of variation in VEVs 
and the structure of the matrices. The inclusion of the singlet Higgs 
field, which enters in the fourth row, is not ad-hoc. Actually we need it 
also to stabilize the supersymmetric $\mu~H_1~H_2$ term\cite{mu} from the 
VEV of the singlet from the trilinear interaction $H_s~H_2~H_1$. However, 
the zero in the $(4,4)$ element is needs to be justified by some 
discrete symmetry. In the case $<H_s>=<H_2>$ we get back two sets of 
degenerate eigenvalues of the Dirac mass matrix. The Majorana mass term 
of the right handed sterile neutrino breaks the $B-L$ symmetry by two 
units. We expect that its mass is tied to the $B-L$ symmetry breaking 
scale $M_R$. 
\section{Conclusions}
In conclusion, we have postulated a set of textures for the 
neutrino Dirac and Majorana mass matrices including a sterile neutrino  
$\n_s$ with one parameter. The Dirac texture is antisymmetric in the 
generation space whereas the Majorana texture is diagonal at 
the leading order displayed in Eqns (\ref{e7}) and (\ref{e8}). We have 
calculated the light neutrino masses upon see-saw mechanism. Solar and 
atmospheric neutrino anomalies can be described by $\n_e \leftrightarrow 
\n_\tau$ and $\n_\mu \leftrightarrow \n_s$ oscillations, whereas the LSND 
results can be described by the $\n_e \leftrightarrow \n_\mu$ 
oscillations. A pair of neutrinos emerge at 0.5-1 eV range whereas 
another pair remain at around $10^{-2}$ eV. It is possible that the heavier 
pair is in suitable mass range for dark matter for the 
$\Omega \approx 0.4$, $\Omega_\Lambda \approx 0.6$\cite{primack}.
The structure of the textures provides large mixing 
angles for $\{e \tau\}$, $\{e \mu\}$ and $\{\mu s\}$ sectors. The right 
handed symmetry breaking scale $V_R$ is bounded from below from the upper 
bound on the left handed Majorana mass from neutrinoless double beta 
decay, whereas it is also bounded from above from the requirement of mass 
difference squared ( $>$ 0.3 eV$^2$) by the LSND results. Hence the scenario 
can be experimentally tested or ruled out in near future. 

\section{Acknowledgements}
\noindent I am grateful to Prof. R. N. Mohapatra for suggestions and 
discussions during this work and to Prof. Joel Primack for 
communications, and Dr. M. Bastero-Gil for comments.

\begin{table}
\begin{center}
\[
\begin{array}{|c|c|c||c|c|c|c|c|c|}
\hline
V_R & \tan \beta  & Log_{10}[\eta] & Log_{10}[\Delta m^2_{e \tau}]&
Log[\Delta m^2_{\mu s}] & \Delta m^2_{e \mu}&
\overline{(\n^e_L)^c}~\n^e_L & m_{\n_e}~(eV) & m_{\n_\mu}~(eV)\\
\hline
10^{14.2} & 2  & -2.0 & -5.08 & -2.03 & 0.93 & 0.49 & 0.03 & 0.97  \\
10^{14.2} & 2  & -2.5 & -5.59 & -2.53 & 0.93 & 0.49 & 0.03 & 0.96  \\
10^{14.2} & 3  & -2.5 & -5.49 & -2.42 & 1.17 & 0.56 & 0.03 & 1.08  \\
10^{14.5} & 2  & -1.5 & -5.17 & -2.12 & 0.24 & 0.25 & 0.01 & 0.50  \\
10^{14.5} & 2  & -2.0 & -5.68 & -2.62 & 0.23 & 0.25 & 0.01 & 0.48  \\
10^{14.5} & 3  & -2.0 & -5.58 & -2.52 & 0.29 & 0.28 & 0.02 & 0.55  \\
\hline
\end{array}
\]
\end{center}
\caption{ For ranges of the parameters in the first three columns, the 
predictions
are given in the last five columns. A crucial prediction in the entry
$\overline{(\n^e_L)^c}~\n^e_L$ of the mass matrix, which is accessible by
present day experiments apart from neutrino oscillations.
}
\label{table3}
\end{table}

\end{document}